\documentclass[proceedings, preprint]{rmaa}

\usepackage{paralist}

\usepackage{psfrag,color}

\SetYear{2020}
\SetConfTitle{Galaxy Evolution: Theory and Observations}

\title{The GWAC Data Processing and Management System} 

\author{
  Y. Xu,\altaffilmark{1,2} 
  L. P. Xin,\altaffilmark{1} 
  X. H. Han,\altaffilmark{1} 
  H. B. Cai,\altaffilmark{1} 
  L. Huang,\altaffilmark{1} 
  H. L. Li,\altaffilmark{1} 
  X. M. Lu,\altaffilmark{1} 
  Y. L. Qiu,\altaffilmark{1} 
  C. Wu,\altaffilmark{1} 
  G. W. Li,\altaffilmark{1} 
  J. Wang,\altaffilmark{3,1} 
  J. Y. Wei,\altaffilmark{1,2} 
  and M. H. Huang\altaffilmark{1,2}}

\altaffiltext{1}{CAS Key Laboratory of Space Astronomy and Technology, National Astronomical Observatories, Chinese Academy of Sciences, Beijing, 100012, China (yxu@nao.ac.cn).}

\altaffiltext{2}{University of Chinese Academy of Sciences, Beijing, 100049, China.}

\altaffiltext{3}{Guangxi Key Laboratory for Relativistic Astrophysics, School of Physical Science and Technology, Guangxi University, Nanning 530004, Peoples Republic of China.}

\shortauthor{Xu, Xin, \& Han, et al.}
\shorttitle{GPMS}

\listofauthors{Y. Xu, L. P. Xin, X. H. Han, H. B. Cai, L. Huang, H. L. Li, X. M. Lu, Y. L. Qiu, C. Wu, G. W. Li, J. Wang, J. Y. Wei, \& M. H. Huang}
\indexauthor{Xu, Y.}
\indexauthor{Xin, L. P. }
\indexauthor{Han, X. H. }
\indexauthor{Cai, H. B. }
\indexauthor{Huang, L. }
\indexauthor{Li, H. L. }
\indexauthor{Lu, X. M. }
\indexauthor{Qiu, Y. L. }
\indexauthor{Wu, C. }
\indexauthor{Li, G. W. }
\indexauthor{Wang,J. }
\indexauthor{Wei,J. Y. }
\indexauthor{Huang, M. H.}

\abstract{GWAC will have been built an integrated FOV of 5,000 $degree^2$ and have already built 1,800 square $degree^2$. The limit magnitude of a 10-second exposure image in the moonless night is 16R. In each observation night, GWAC produces about 0.7TB of raw data, and the data processing pipeline generates millions of single frame alerts. We describe the GWAC Data Processing and Management System (GPMS), including hardware architecture, database, detection-filtering-validation of transient candidates, data archiving, and user interfaces for the check of transient and the monitor of the system. GPMS combines general technology and software in astronomy and computer field, and use some advanced technologies such as deep learning. Practical results show that GPMS can fully meet the scientific data processing requirement of GWAC. It can online accomplish the detection, filtering and validation of millions of transient candidates, and feedback the final results to the astronomer in real-time. During the observation from October of 2018 to December of 2019, we have already found 102 transients.}

\resumen{GWAC tiene como objetivo un campo de visión integrado de 5000 grados$^2$ de los cuales ya se han construido 1800 grados$^2$.
La magnitud limite de una imagen de exposición de 10 segundos en una noche sin luna es 16. Asi en cada noche de observación, GWAC produce aproximadamente 0.7 TB de datos sin procesar. Una vez los datos son procesados, se generan millones de alertas de una sola vez. Describimos el procesamiento de datos GWAC y el sistema de gestión (GPMS), que incluye arquitectura de hardware, base de datos, detección-filtrado-validación de candidatos transitorios, archivo de datos e interfaces de usuario para la verificación de transitorios y monitorización del sistema. GPMS combina tecnologia general y software en astronomia, utilizando algunas tecnologias avanzadas como el aprendizaje profundo. Los resultados prácticos muestran que GPMS puede cumplir con los requisitos cientificos de procesamiento de datos de GWAC. Por ello se puede realizar en linea la detección, filtrado y validación de millones de candidatos transitorios, y retroalimentar con los resultados finales al astrónomo en tiempo real. 
Durante la observación de octubre de 2018 a diciembre de 2019 se detectaron 102 transitorios.

}

\addkeyword{astronomical databases: miscellaneous}
\addkeyword{catalogs}
\addkeyword{methods: data analysis }
\addkeyword{techniques: image processing}
\addkeyword{techniques: photometric}

\begin{document}
\maketitle

\section{Introduction}
\label{sec:intro}

The Ground-based Wide-Angle Cameras array (GWAC) is a ground-based observation device for the Sino-French astronomical satellite SVOM mission \citep{2015arXiv151203323C, 2016arXiv161006892W}. Its main scientific goal is to discover short-time transients based on the large field-of-view (FoV) optical survey, to make up the gaps of real-time detection of transients with the time-scale of hours or shorter. GWAC has the advantages of large FoV and high observation cadence and can monitor the sky of 5000 square degrees in real-time. The high cadence monitoring feature enables the ability to detect flared events with a very short time scale, so the data processing system is required to complete the detection and filtering of the transient in real-time, and identify the transient in time. 

Some science results and technical details about GWAC are as follows.  The super outbursts of three dwarf novae identified independently by GWAC are reported by \citet{2020AJ....159...35W}, including the follow-up observations of photometric and spectroscopic. \citet{2019arXiv190208476T} describe the optical follow-up results of the second observational campaign of gravitational waves with mini-GWAC that is the pathfinder of GWAC. \citet{2016PASP..128k4501W}  designed a distributed database to manage catalogs observed by GWAC with more than 100 billion records for 10 years of observation. \citet{2015Huang} introduced an auto-focusing system based on step motor and full width at half maximum (FWHM) of image for wide-angle telescopes, which greatly improves the quality of the image. 

This paper gives a high-level overview of the GWAC Data Processing and Management System (hereafter GPMS) housed at the dome of GWAC. Core functions of GPMS include the detection, filtering, validation of transient candidates; real-time alert for transients and emergency of the system; user interfaces for the check of transient and the monitor of the system; long-term archiving and history image retrieval. Section 2 introduced the GPMS’s hardware architecture. The pipelines and tools are described in Section 3, and some statistical results are introduced in Section 4. Section 5 summarizes this paper.

\section{Hardware architecture}
\label{sec:hwArch}

The hardware architecture of GPMS is shown in Figure 1, including the GWAC telescopes, follow-up observation telescopes, control computer, data processing computer, Web server, database server, and the disk array. GWAC plans to build a telescope array with a total FoV of about 5,000 square degrees and the limit magnitude of an image with a 10-second exposure in the moonless night is about 16.0R. Three units of GWAC have been completed, covering almost 1,800 square degrees of FoV. 

Each unit contains four cameras with a diameter of 18 cm and one camera with a diameter of 7 cm. The main function of the 7 cm telescope is for the direction of the mount, and it can also supplement the observation of bright stars saturated on the 18 cm telescope. Each camera is equipped with a control computer and a data processing computer. The control computer is located in the dome near the telescope, used to control the CCD, collect images and monitor the hardware parameters such as vacuum, voltage and current in real-time. The data processing machine is a high-performance server, which is located in the computer room and is used to complete data processing tasks such as transient source detection and filtering, and scientific product upload. 

\begin{figure*}[!t]
  \includegraphics[width=2\columnwidth]{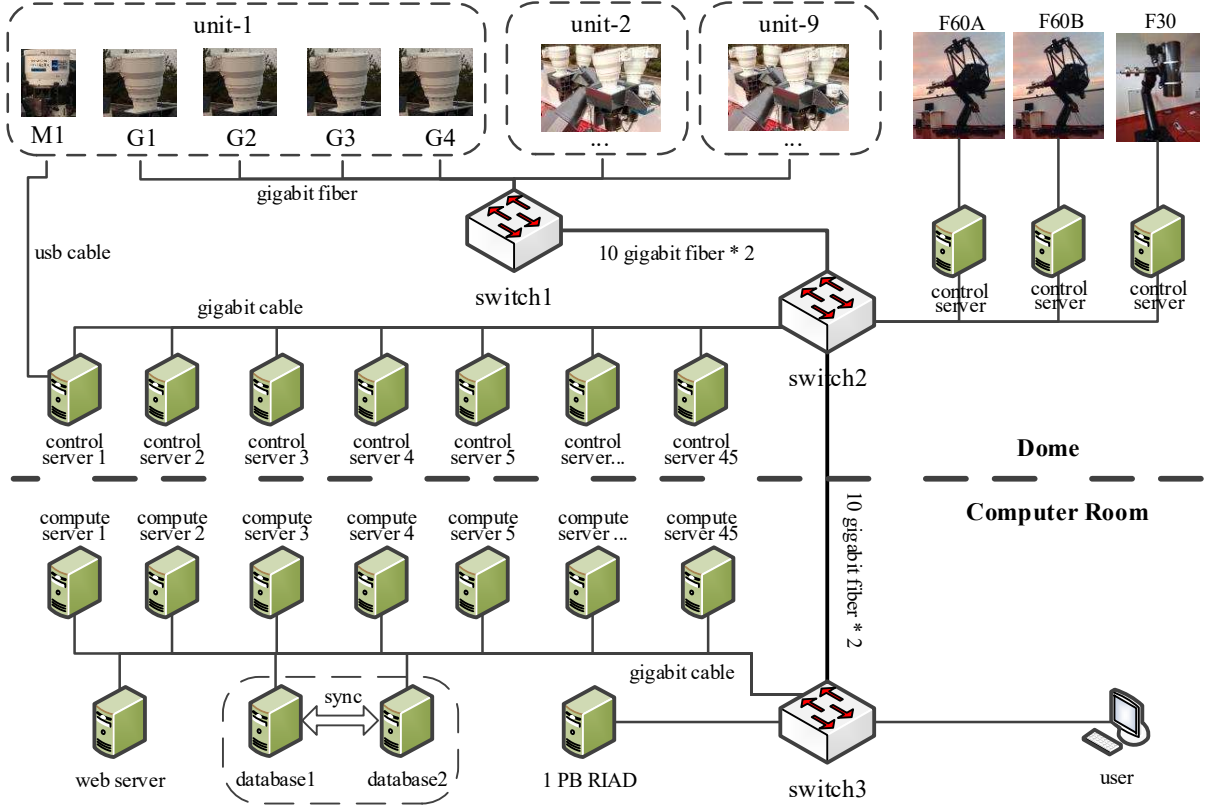}
  \caption{The hardware architecture of GPMS, including the GWAC telescopes, follow-up observation telescopes, a computer cluster, and a disk array.}
  \label{fig:architecture}
\end{figure*}

The output of 18 cm cameras is a fiber interface. Its data are converged in a fiber switch and transmitted to the data processing computer through two 10-Gigabit fiber network. The output of 7 cm cameras are USB interface and are directly connected to the corresponding CCD control computer. In order to simplify the data processing logic and management, we adopt a one-to-one method, that is, each CCD is controlled and monitored by a control computer, and also has a data processing computer. 

The web server is used to collect all the scientific products produced by the data computer and provides a user interface for displaying the list of transients and their detailed information. The database stores information such as catalogs, image quality, instrument status, etc. The 1PB disk array is used to archive all observation images and scientific products. There are also two 60 cm telescopes (F60s) and one 30 cm telescope, which are mainly used for the follow-up observation of GWAC transient candidates, and also participate in the follow-up observation of gravitational waves, GRBs, etc.

\section{Pipelines and tools}
\label{sec:pipelines}

Figure 2 shows the core data processing flow of GPMS, including detection, filtering and verification of transients, and sending of alerts. In addition, GPMS also includes automatic light curve observation of validated transient, data archiving and query, and user support interface. The detection of transients includes two methods: catalog matching and image subtraction. The transient candidates are filtered by known star catalog matching, morphological parameters, and machine learning, and then verified by the follow-up of F60s. Finally, an alert message is sent for each verified transient to inform the scientist in time.

\begin{figure*}[!t]
  \includegraphics[width=2\columnwidth]{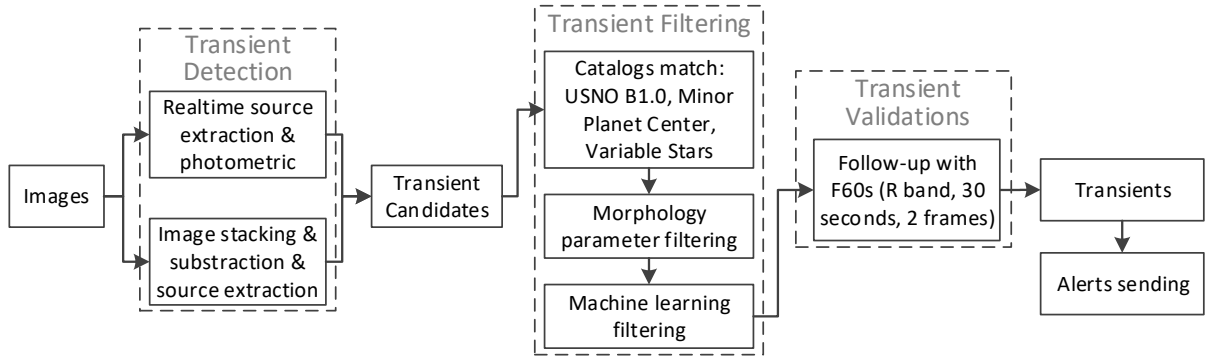}
  \caption{The core data processing flow, including detection, filtering and verification of transients.}
  \label{fig:pipeline}
\end{figure*}

\subsection{Transient detection}
\label{sec:detection}

Transient detection includes two pipelines: catalog matching and image subtraction. The catalog matching is faster, which can meet the real-time transient detection needs of GWAC. Image subtraction can effectively deal with dense star fields and galaxies. The control computer cache the image in local disk after the CCD accomplishes exposure and then transmits the image to the data processing computer through the 10 Gigabit fiber network, and the transient detection pipelines will immediately process those images.

The pipeline of catalog matching is the main transient detection process of GWAC, and it has been running in real-time for many years. The data processing process of this pipeline mainly includes template making, point source extraction, cross-match, astrometry and photometry, and other general data processing procedures. The template image is produced by selecting and combining 10 high-quality images from historical observation images, and the template catalog is extracted from template image; if the current observation sky area has not been observed previously or the historical image is not good, the new observation image is used to make the template. Point source extraction is performed by calling Sextractor \citep{1996A&AS..117..393B}. The cross-match first performs a position comparison match with a radius of 1 pixel in the image coordinate; and if the match fails, a triangular blind match is performed. About astrometry, we found that Astrometry.net  \citep{2009PhDT.......235L} has large errors in some sky areas, so our project team has built a custom astrometry index for GWAC.

The image subtraction pipeline is currently in the trial operation phase, and its data processing process is similar to the catalog matching, but adding the image combination and subtraction processes. Image combination includes the series of 5, 25, and 125 frames of stack. The current subtraction pipeline performed by calling hotpants \citep{2015ascl.soft04004B} that has high compatibility and can be steadily used for the subtraction of GWAC's large FoV and under-sampling images. We have also tested ZOGY \citep{2016ApJ...830...27Z} needed an engineering adaptation for large FoV.

\subsection{Transient filtering}
\label{sec:filtering}
The pipelines of catalog matching and image subtraction are found targets that are not in the template, these targets are defined as transient candidates. The transient candidates include variable stars, moving targets, artificial targets, etc. With the multi-frame appearance strategy, most fast-moving targets can be filtered. Filtering methods also include known star catalog matching, morphological parameters, and machine learning.
\begin{asparaitem}
\item Filtering by matching known star catalog: Each transient candidate is matched with known star catalogs such as Minor Planet Center, variable star catalogs, USNO B1.0, 2mass, Gaia Dr2, etc., which can filter asteroids, variable stars, and some residuals targets caused by image subtraction.
\item Filtering by morphological parameters: The ellipticity, FWHM, and other morphological parameters can be used to filter fast-moving targets, hot pixels, some CCD defects, etc.
\item Filtering by Machine learning: The above two methods cannot effectively filter false transient sources such as CCD defects, dust, and residuals targets caused by image subtraction. So we proposed the multi-size convolutional neural network \citep{xu2020a} for the classification of optical transient, which greatly reduces the number of false transients.
\end{asparaitem}

\subsection{Transient validation}
\label{sec:validation}
After the filtering in section 3.2, most false transients are filtered out. However, there still contain some false transients, which need to be validated by follow-ups. For each target, GPMS automatically calls the F60s to continuously observe two R-band frames with 30 seconds exposure. According to the observation results, the candidates can be classified into the following three categories:
\begin{asparaitem}
\item Artificial defects: F60s do not detect any objects at 5 sigma significance level within 30 arc-seconds around the candidate positions determined by GWAC. 
\item Flares or large-amplitude variables: A candidate is classified as either a flare star or a large-amplitude variable in the following conditions. At first, F60s detect at least one source within the same area. Secondly, each source has a counterpart within 3 arcsec in known catalogs, such as USNO B1.0, 2mass, Gaia Dr2. Finally one of them is brighter by more than one magnitude compared to the catalogs.
\item New transient: Some of the source detected by F60s within 30 arcseconds around the candidate positions determined by GWAC have no counterpart within 3 arcsec in the known catalogs.
\end{asparaitem}

\subsection{Automatic light-curve sampling for confirmed transient}
\label{sec:sampling}
Once a transient is confirmed, an automatic and flexible exposure sequence of light curve sampling taken by F60s is necessary for obtaining maximum scientific returns with minimum costs after taking into account the requirements of the balance of multi-tasks and the balance of exposure time and cadence. So it is necessary to dynamically adjust the exposure time, filter and cadence based on the evolution of multi-band light curves, we proposed a real-time automatic validation system, more detail can be seen in \citep{xu2020b}.

\subsection{Real-time Notification}
\label{sec:notification}
The scientist or instrument engineer needs to be notified immediately when a transient candidate is validated or an emergency occurred. GPMS uses enterprise WeChat\footnote{https://work.weixin.qq.com} as a solution of the real-time mobile notification for pushing validation results of each transient candidates and ancillary data. Enterprise WeChat provides a wealth of user and message management functions to support the transmission of text, images, files, voice, video and other types of information. The message sending interface is encapsulated as a web service, which is convenient for the call by all pipelines. 

\subsection{Long-term archiving and retrieval}
\label{sec:archiving}
On the day of observation completion, the observation images are not backed up. Instead, those images are lossless compressed by fpack \citep{2010ascl.soft10002S} and centralized backup to the 1PB disk array and organized by date and camera name in the next day.

An image is selected as a representative for each sky after each day’s observation. Its center coordinate is calculated and is used to establish a joint index with observation time. Then we can query historical observation records and image lists based on location and time.

\subsection{User interfaces for transient checking and system monitoring}
\label{sec:interfaces}
We built a web information system based on the Java platform to receive and display the transients generated by the data processing pipeline in real-time. The functions include user login and management, manual follow-up, the display of transients list and detailed information including cutouts, light curves from GWAC and F60s, position changes, catalogs matching results.

\section{Statistic}
\label{sec:statistic}
At present, the main detection pipeline of GPMS is the catalog matching pipeline. We counted the total time of the recognition and filtering process of the 10655 transient candidates that are detected by catalog matching and automatically validated by follow-up of F60s. The statistical results are shown in Figure 3. The minimum value is 5 seconds, that is, it takes only 5 seconds to complete the recognition and filtering of the transient candidates, and to start the follow-up observation of the transient. The average is 25.1 seconds, and only 9.1\% is greater than 60 seconds.

\begin{figure*}[!t]
  \includegraphics[width=2\columnwidth]{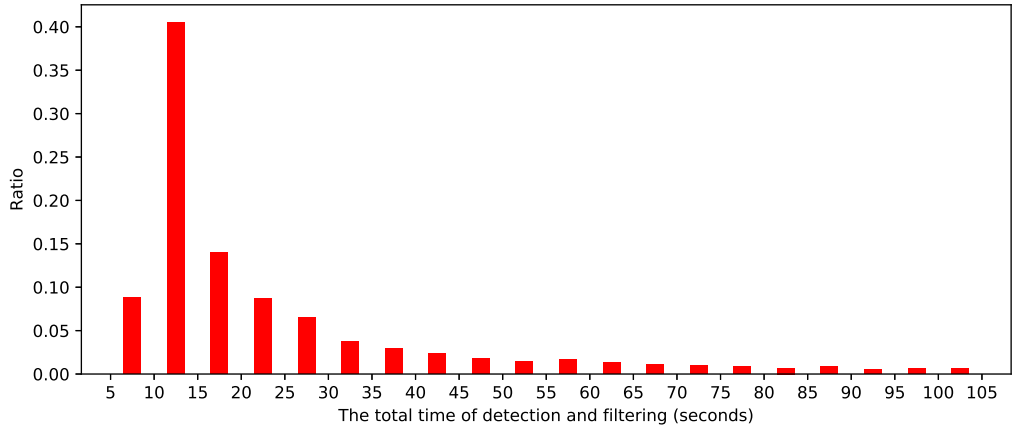}
  \caption{Total time statistic of the recognition and filtering process of the 10655 transient sources that are automatically validated by follow-up. The minimum value is 5 seconds, the median value is 15.3 seconds, and the average value is 25.1 seconds.}
  \label{fig:AutoTriggerTime}
\end{figure*}

GWAC officially started operation in October 2018, and the single-frame catalog matching pipeline has detected 102 transient source targets by December 2019. The R-band magnitude and magnitude difference of the first detected frame of 79 transients with complete identification data are counted in figure 4. Left is the first frame detection magnitude statistics of the transient, the faintest is 14.9. Right shows the statistic of magnitude difference with the matched known star of the transient’s first detected frame. The average of the magnitude difference is 3.6, and only 16.5\% of the magnitude of the change is less than 2.

\begin{figure*}[!t]
  \includegraphics[width=2\columnwidth]{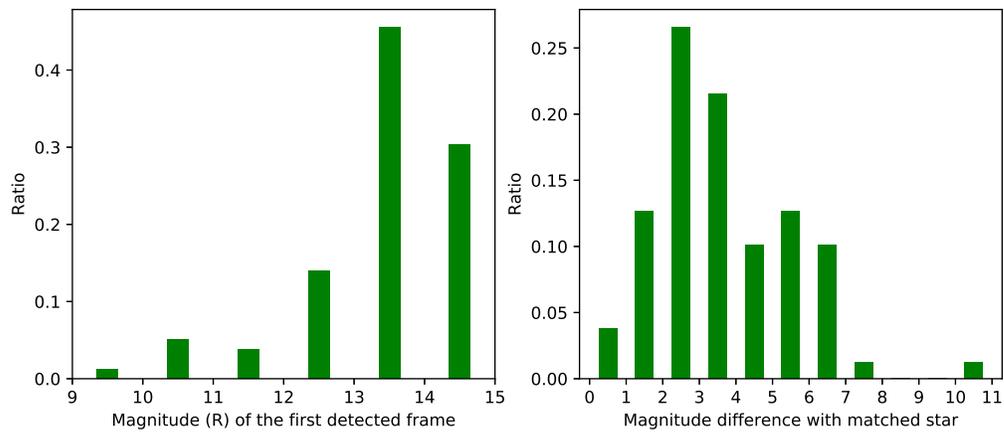}
  \caption{Single-frame detection result statistic of catalog matching pipeline. The R-band magnitude (left) and magnitude difference (right) of the first detected frame of 79 transients with complete identification data.}
  \label{fig:magnitudeStatistic}
\end{figure*}

We will gradually optimize the data processing flow of GWAC in the future. For example, image combination can extend the detection limit to 18.0R. Training efficient machine learning algorithms can improve the detection rate of low-amplitude targets.

\section{Summary}
\label{sec:summary}
In order to support the massive data processing and real-time transient detection requirements of GWAC, we designed the GPMS. The hardware architecture of GPMS is briefly introduced, including hardware composition, network connection, data processing and exchange methods. At the same time, we also introduced the GPMS data processing process in detail, including detection, filtering and verification of transient, automatic follow-up of light-curve, real-time message notification, data archiving and query, user interface of transient checking and system monitoring. 
The minimum time consumed by the recognition and filtering process of GPMS is  5 seconds.
During the observation from October of 2018 to December of 2019, we have already found 102 transients.
Among these transients, 79 transients with complete identification result are counted, the faintest is 14.9R, and 83.5\% magnitude change is greater than 2.

\end{document}